\documentclass[fleqn,usenatbib]{mnras}

\usepackage{newtxtext,newtxmath}
\usepackage{newtxtext,newtxmath}

\usepackage[T1]{fontenc}

\DeclareRobustCommand{\VAN}[3]{#2}
\let\VANthebibliography\thebibliography
\def\thebibliography{\DeclareRobustCommand{\VAN}[3]{##3}\VANthebibliography}

\usepackage{graphicx}	
\usepackage{amsmath}	
\usepackage[normalem]{ulem}

\def\specchar#1{{\sc{#1}}}    
\def\specand{\,\&\,}               
\def\Halpha{\mbox{H\hspace{0.1ex}$\alpha$}}
\def\CaII{\mbox{Ca\,\specchar{ii}}}
\def\CaIR{\mbox{Ca\,\specchar{ii}\,\,8542\,\AA}}
\def\CaIIK{\mbox{Ca\,\specchar{ii}\,\,K}}
\def\CaIIHK{\mbox{Ca\,\specchar{ii}\,\,H{\specand}K}}
\def\Kthree{\mbox{K$_3$}} 
\def\Ktwo{\mbox{K$_2$}}
\def\KtwoV{\mbox{K$_{2V}$}}
\def\KtwoR{\mbox{K$_{2R}$}}
\def\ktwoR{\mbox{k$_{2R}$}}
\def\kthree{\mbox{k$_{3}$}}
\def\ktwo{\mbox{k$_{2}$}}
\def\MgII{\mbox{Mg\,\specchar{ii}}}
\def\MgIIhk{\mbox{Mg\,\specchar{ii}\,\,h{\specand}k}}

\def\MgIIk{\mbox{Mg\,\specchar{ii}\,\,k}}
\def\SiIV{\mbox{Si\,\specchar{iv}}}

\def\CII{\mbox{C\,\specchar{ii}}}
\def\SST{Swedish 1-m Solar Telescope}
\def\SDO{{\it Solar Dynamics Observatory\/}}
\def\IRIS{{\it Interface Region Imaging Spectrograph\/}}
\def\CRISP{CRisp Imaging SpectroPolarimeter}

\def\EBs{Ellerman bombs}
\def\UVBs{UV bursts}
\newcommand{\xy}[2]{{{($X$,$Y$) = (#1\arcsec,#2\arcsec)}}}

\title[COCOPLOT: COlor COllapsed PLOTting software]{COCOPLOT: COlor COllapsed PLOTting software \\ Using color to view 3D data as a 2D image}

   \author[M. K. Druett et al.]{Malcolm K.~Druett$^{1,2}$\thanks{E-mail: malcolm.druett@kuleuven.be (SU)},
          Alexander G.~M.~Pietrow$^{1,3}$,
          Gregal J.~M.~Vissers$^{1}$,
          Carolina Robustini$^{1}$, 
          Flavio Calvo$^{1}$
\\
$^{1}$Institute for Solar Physics, Department of Astronomy, Stockholm University, AlbaNova University center, SE-106 91 Stockholm, Sweden
\\
$^{2}$Centre for mathematical Plasma Astrophysics, Department of Mathematics, KU Leuven, Celestijnenlaan 200B, B-3001 Leuven, Belgium
\\
$^{3}$Leibniz-Institut für Astrophysik Potsdam (AIP), An der Sternwarte 16, 14482 Potsdam, Germany
}
\date{Accepted XXX. Received \today ; in original form ZZZ}

\pubyear{2015}

\begin{document}
\label{firstpage}
\pagerange{\pageref{firstpage}--\pageref{lastpage}}
\maketitle

\begin{abstract}
Most modern solar observatories deliver data products formatted as 3D spatio-temporal data cubes, that contain additional, higher dimensions with spectral and/or polarimetric information. This multi-dimensional complexity presents a major challenge when browsing for features of interest in several dimensions simultaneously. We developed the COlor COllapsed PLOTting (COCOPLOT) software as a quick-look and context image software, to convey spectral profile or time evolution from all the spatial pixels ($x,y$) in a 3D [$n_x,n_y,n_\lambda$] or [$n_x,n_y,n_t$] data cube as a single image, using color. This can avoid the need to scan through many wavelengths, creating difference and composite images when searching for signals satisfying multiple criteria. Filters are generated for the red, green, and blue channels by selecting values of interest to highlight in each channel, and their weightings. These filters are combined with the data cube over the third dimension axis to produce an $n_x \times n_y \times 3$ cube displayed as one true color image. Some use cases are presented for data from the \SST\ and \IRIS\ (IRIS), including \Halpha\ solar flare data, a comparison with $k$-means clustering for identifying asymmetries in the \CaIIK\ line and off-limb coronal rain in IRIS \CII\ slit-jaw images. These illustrate identification by color alone using  COCOPLOT of locations including line wing or central enhancement, broadening, wing absorption, and sites with intermittent flows or time-persistent features. COCOPLOT is publicly available in both IDL and Python.
\end{abstract}




\section{Introduction} \label{sec:intr}
The view of the solar atmosphere can vary greatly even within the wavelength range of a single spectral line due to its wavelength-dependent opacity, with photons of different wavelength escaping from different height ranges.
While this allows sampling several atmospheric layers by scanning through a spectral line, one is often also interested in looking for Doppler-shifted emission or absorption, or effects like line broadening or enhancement of the wing(s) or core.
Therefore, a major challenge in analyzing observations is to access the data in a way that enables differentiating between intensities in several regions of the spectral line simultaneously, preferably for many pixels at the same time. Doing so allows selecting features of interest for further study or highlighting them within the field-of-view for presentation. 

Many quick-look packages offer flexible browsing functionality to explore the different dimensions encapsulated in the data (e.g.~CRISPEX \citep{2012VissersCrispex, 2018arXiv180403030L}, SOAImage DS9 \citep{2003DS9} or IRAF \citep{tody93}), but currently the user is left to scan and blink between monochromatic images of specific wavelengths or to generate gray-scale composites and difference images. 
While this works for simple features that are identifiable using `net emission' or intensity differences between two spectral positions, this is more challenging when a greater level of spectral information must be conveyed in order to analyze the atmosphere. For example, the double peaked shape of the  \CaIIHK\ lines makes it difficult to locate asymmetries without manually searching through image pixels. 

In night-time astronomy several methods are already commonly used that encode multiple features into color images. The most simple of these assigns three separate wavelength filters into the three channels of an RGB image \citep{Malin1992}. In order to view detail in separate regions with very different intensities, more complex variations exist that allow the user to scale the RGB values against the logarithm of intensity, along a square root, or for instance with a Lupton filter \citep{lupton2004, Villard2002}.

\begin{figure*}
\centering
   \includegraphics[width=0.99\textwidth]{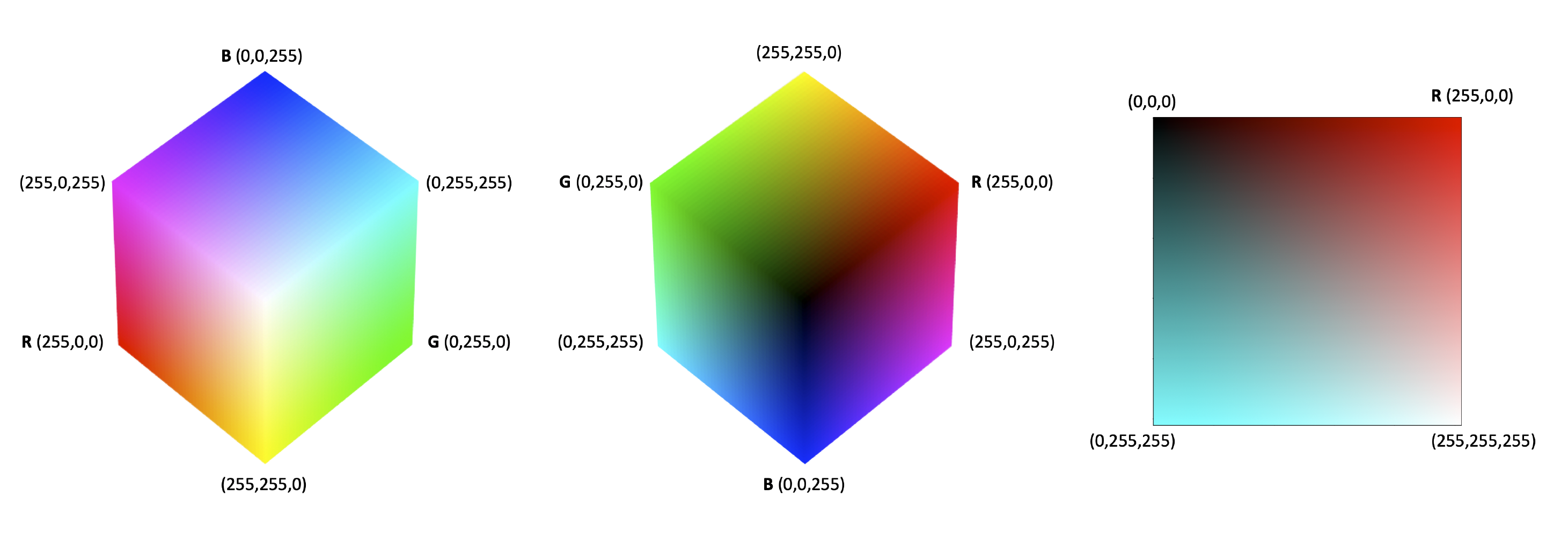}
   \label{fig:rgbcube}
\caption{A representation of the RGB color space where all possible RGB colors are a function of the three primary colors. On the left side we see the RGB cube as seen while looking at the white corner, which represents the RGB value of (255,255,255). The central figure shows the same cube but from the back, looking at the black corner, which represents the RGB value of (0,0,0). The right figure shows a diagonal cut through the RGB cube to show the gray scale colors on the diagonal from top left to bottom right, which are obtained when all three RGB values have the same number.}
\label{fig:colorwheel}
\end{figure*}

In some applications the RGB channels are not necessarily the optimal choices. The Noise Adaptive Fuzzy Equalization (NAFE) algorithm \citep{2013NAFE} was developed to display small features without saturating the image when the data source has an extremely large range, namely AIA-SDO data \citep{AIA}. A Planckian Mapping evolution of this algorithm (PM-NAFE) combines the three AIA wavelength bands with the most single-peaked temperature response function using colours that allow the reader to intuitively interpret characteristic temperature maps. The colours chosen for each band relate to the characteristic colours of emissions from black-bodies to help the reader intuitively infer characteristic temperatures. the leads to a visual impression of cold dark regions, through reddish hues, to white-hot scale. 

Approaches have been employed in solar physics to visualize e.g. 
rapid blue-shifted events in \Halpha\ and \CaIR\ 
\citep{2012ApJ...752..108S, 2013ApJ...764..164S}, 
off-limb coronal rain in coordinated observations between the \SST\ (SST),
\SDO, \IRIS\ (IRIS) and Hinode
\citep{2015ApJ...806...81A}, 
penumbral microjets in IRIS and SST observations \citep{2015ApJ...811L..33V}. 
However, these methods are either designed for single wavelength filters that do not require more specific spectral profile information or were applied to particular spectral line positions. 
As solar observations often provide spectral line sampling at a large number of wavelength positions for extended fields-of-view, such methods cannot be readily applied to condense a larger range of spectral information in a single image.

To address this challenge we developed the `COlor COllapsed PLOTting' (COCOPLOT) software. Its underlying procedure draws inspiration from the way that our eyes interpret complicated spectra by utilizing three different types of optical receptors, with each receptor having a distribution of sensitivity across the visible light spectrum. We translate this approach to solar data by defining three filters that can be applied to any 3D data cube, by collapsing its third dimension with the filters, in order to create a weighted `white light image'. Using the standard RGB table with values from 0 to 255 over 16 million color hues can be created as a combination of the three primary colors, allowing the user to convey the huge variation in the data present across the collapsed dimension of the data cube. This method can therefore effectively summarize the spectral or temporal information across the data cube in a single quick-look or context image.

\begin{figure*}
\includegraphics[width=\textwidth]{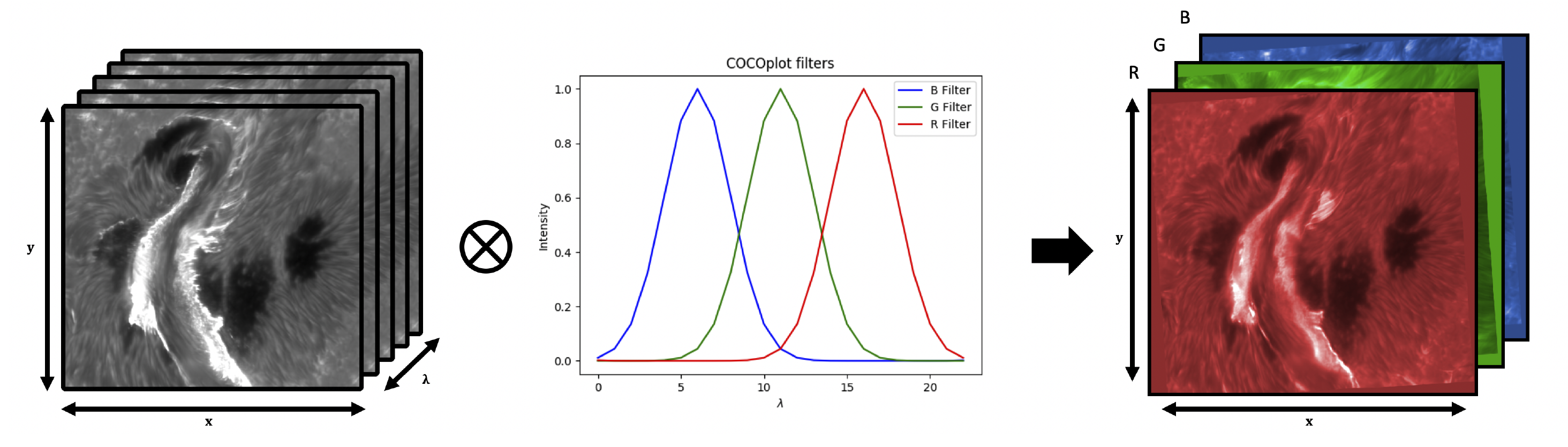}
\caption{Cartoon showing the basic COCOPLOT method, whereby a 3D data cube (consisting of 2D monochromatic images) is collapsed using three separate filters that have the same length as the spectral axis of the data cube. Each combination of a filter with the cube defines a channel in the resulting RGB image that can in turn be displayed as a single true color image.}
\label{fig:filter}
\end{figure*}
Importantly, COCOPLOT provides the freedom to place the filters anywhere in the spectrum and to adjust their form and sensitivity. Thus one can easily set up COCOPLOTs that react strongly to certain features while ignoring others, allowing the user to identify and highlight regions of interest in their data using the differences in their spectra. Moreover, this method is not limited to spectral data, as it could also be applied to any 3D cube when appropriate filters are chosen.
This software is made freely available in Python and IDL, and is written with the intent to be easy to use alone, or use alongside or within existing quick-look tools. The software will also be callable as an extension in an upcoming update of CRISPEX.

The remainder of this paper is structured as follows. In section \ref{sec:use} we describe the COCOPLOT software and how to interpret color in the images. In section \ref{sec:results} a number of science uses are discussed. COCOPLOTs are used on spectral data obtained from the SST and IRIS in sections \ref{sec:SST} and \ref{sec:IRIS} respectively. A use of COCOPLOTs with non-spectral information are presented in \ref{sec:time}, in which we summarize time-series data using the same method. In section \ref{sec:conclusions} the value of the method for science uses as well as its limitations are discussed. 

\section{Using COCOPLOTs} \label{sec:use}

\subsection{COCOPLOT method description} \label{sec:code}
The COCOPLOT software suite consists of a set of functions and procedures that allow the user to create and apply RGB filters, as well as visualize and save the result, be it as single images or as an animated sequence of COCOPLOTs.
In this section we describe the process by which a COCOPLOT is generated using the example of collapsing the wavelength dimension of a spectral data cube (e.g. a single wavelength scan from an imaging spectropolarimeter).
We note, however, that the third dimension of the input data cube may also be e.g. time (or height in a simulation data cube) and hence in the following spectrum/spectral could be read interchangeably as time/temporal (or height).

In order to construct a COCOPLOT image, the user must supply a three-dimensional data cube ($n_x, n_y, n_{\lambda}$) and three filters for the red, green, and blue channels of the color image (Fig.~\ref{fig:filter}). The individual filters are 1D arrays of length $n_{\lambda}$ that can be user-defined or may be generated using the {\bf cocofilter} function. The elements in the filter matrix ($n_{\lambda}$,3) are numbers between 0 and 1, defining the desired weights of each wavelength position in the data cube for the red, green, and blue filters. The spectral dimension of the data cube is then multiplied with the filter entries and the total taken for each filter, similar to taking the dot-product of the data cube with filter matrix. This produces an $n_x \times n_y \times 3$ RGB array. This process is visualized in Fig.~\ref{fig:filter}. In this cartoon the red filter is used to cover the red part of the spectrum, which is the larger values for wavelength data shown (Fig.~\ref{fig:filter}), but we note that it should be used to cover the lower values for frequency data. The resulting RGB array can then be displayed as a single "true 24 bit color" image. Further function/procedure details and code examples are provided in Appendix~\ref{app:functions}.

\subsection{Interpreting COCOPLOT colors} \label{sec:colors}

\begin{figure*}
\centering
   \includegraphics[width=\textwidth]{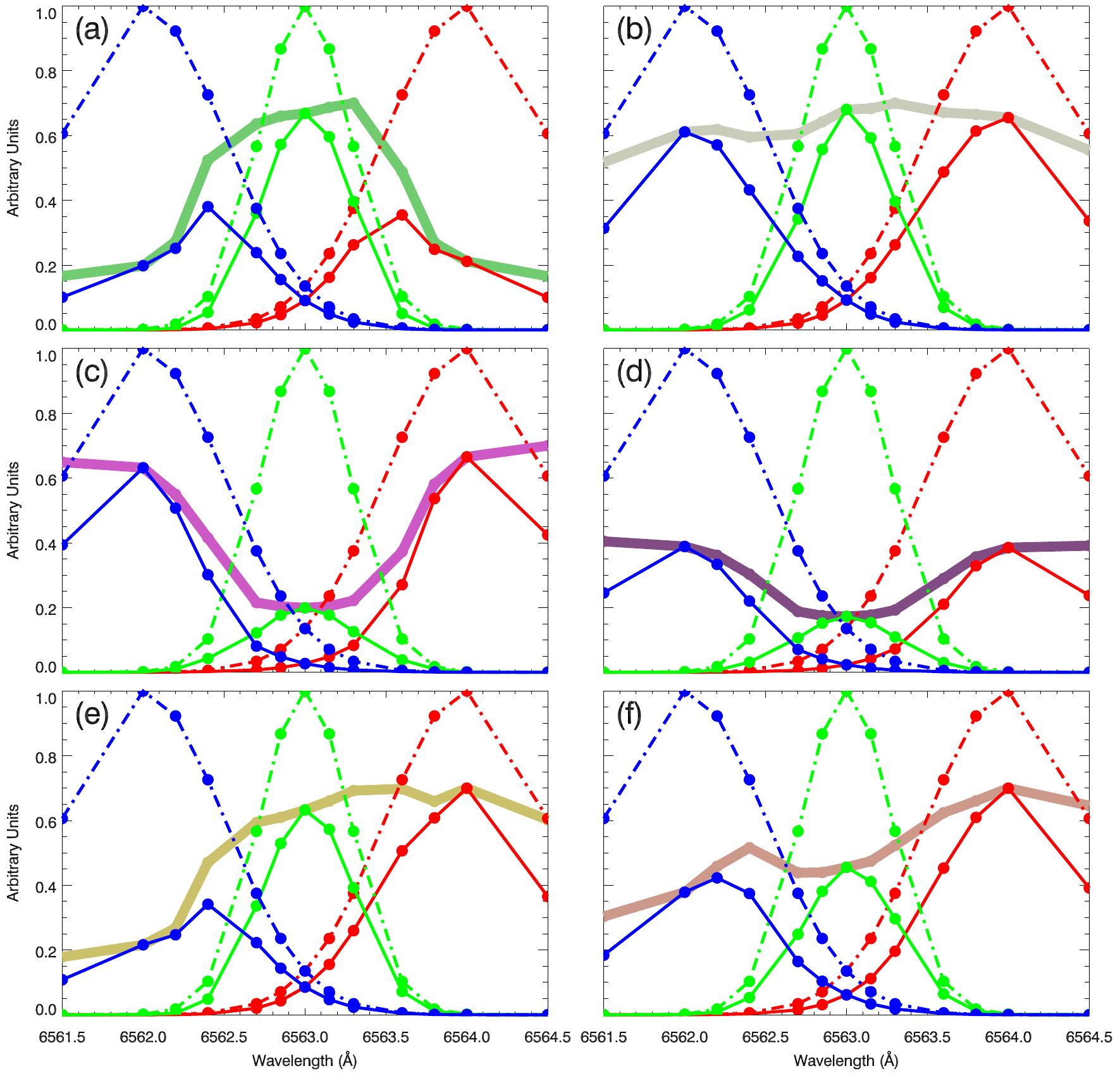}
\caption{Understanding COCOPLOT colors: (a)--(f) examples of how color is used to encode the \Halpha\ 6563\,\AA\ profile shape ({\it solid lines in hues representing COCOPLOT color\/}) through combination of the RGB channels ({\it solid red, green and blue lines\/}) that result from taking the product of the profile by the user-defined filters ({\it dashed red, green and blue lines\/}).
The y-axis values are normalized to arbitrary values convenient for display. 
}
\label{fig:colors}
\end{figure*}

RGB values are represented by a three-element array of integers, each element taking a value between 0 and 255. The colors resulting from combining RGB channels are shown by the  RGB cube in Fig.~\ref{fig:colorwheel}. 
We first discuss the colors encoded by different RGB arrays for a few cases that will prove to be relevant in subsequent discussions of interpreting line profile forms from a COCOPLOT: 

\begin{itemize}
\item One element dominates: If one of the red, green, or blue channel elements is much higher than the other two, then the resulting RGB color will be close to that primary color with the higher value (e.g. such as the green channel in Fig.~\ref{fig:colors}a).
\item Monochrome: If RGB values are low in all three elements then the color will be black. If the values are higher but approximately equal then the color will be grey. The closer these values are to 255, the closer to white the color will be  (Fig.~\ref{fig:colors}b).
\item Two elements dominate: If the R and B elements have much higher values than the G channel, then the resulting color combination is purple or magenta (Fig.~\ref{fig:colors}c). If the R and G elements are both much higher than the B element then the combination produces yellow. Finally, if the B and G elements are both much higher than the R element, the resulting color is cyan.
\item Brightness: If an RGB array has a high valued entry such as [255,0,255], it will produce a bright color (Fig.~\ref{fig:colors}c). If the value is instead lower, for instance [127,0,127], then the result will be a darker color (Fig.~\ref{fig:colors}d).
\item More subtle differences: If the R and G elements are both much higher than the B element then the combination produces yellow (Fig.~\ref{fig:colors}e). However, the result will be an orange-red hue if the highest data values are noticeably in the R element rather than the G (Fig.~\ref{fig:colors}g).
\end{itemize}

The exact nature of the information conveyed by the colors in a COCOPLOT depends on the chosen RGB filters, i.e. the shape, width and position. It is paramount that these filters are selected such that important features of the spectral dimension of the data cube may be understood by considering only the colors provided (see Fig.~\ref{fig:colors}). To illustrate this in further detail, we refer back to the panels in Fig.~\ref{fig:colors} which are actually examples taken from an observations of a solar flare in the hydrogen Balmer $\alpha$ (\Halpha) spectral line at 6563\,\AA. 
The COCOPLOT filters used have a normal distribution with wavelength. They are positioned at the nominal $\Halpha$ line center (green) and out at $\pm$1\,\AA\ (red and blue), with a standard deviation of $\sigma=0.28$\,\AA\ for the green filter and $\sigma=0.5$\,\AA\ for the red and blue filters.
Panels~\ref{fig:colors}a--f show normalized line profiles (solid colored lines), filter values (dashed RGB color lines) and the values produced by combining the filter values with the profiles (solid RGB color lines). The hue of the line profiles that are used are similar to the resulting colors of the pixels in the COCOPLOT at the spatial locations of these profiles. The locations of the pixels where these data are taken from are shown in Fig.~\ref{fig:flarecompare} for reference, using labelled circles with a similar color to the respective profile lines in Fig.~\ref{fig:colors}.

Figure~\ref{fig:colors}c shows an absorption profile with higher wings that cover the wavelengths of the blue and red filters, but low values in the line center. The mixture of mainly blue and red colors, with absence of green, results in a purple pixel color in the COCOPLOT. 
Figure~\ref{fig:colors}a shows a profile that is mainly enhanced at line center. These have high coefficients in the green filter, therefore the profile will be represented by a green hue. 
Figure~\ref{fig:colors}b shows a line profile that has similar values across all wavelengths, resulting in a balanced mix of red, green, and blue. This results in the hue of the pixel being monochrome gray in the COCOPLOT. Because the intensities here were high, the hue is close to white.
Figure~\ref{fig:colors}e shows a profile that is centrally enhanced and also enhanced in the red wing, resulting in a mixture of mainly green and red colors, producing a yellow COCOPLOT pixel. Finally, Fig.~\ref{fig:colors}f shows a similar profile, but with less of a central enhancement and somewhat higher blue wing. As red dominates, the pixel hue is therefore redder by comparison.

\section{Results} \label{sec:results}
In this section we present a number of COCOPLOT use cases, ranging from identification of evolutionary features in \Halpha\ flare observations (section~\ref{sec:Halpha}), providing an alternative to $k$-means clustering of \CaIIK\ profiles (section~\ref{sec:CaK}), and identification of UV bursts in IRIS data (section~\ref{sec:IRIS}) to characterizing the time evolution of flares, coronal rain and flux emergence activity (section~\ref{sec:time}). 
These represent a variety of science use cases, but this list is by no means exhaustive.

\subsection{Visualizing spectral features} \label{sec:SST}
\subsubsection{Evolution of \Halpha\ features in a flare} \label{sec:Halpha}

\begin{figure*}
   \centering
   \includegraphics[width=0.833333333\textwidth]{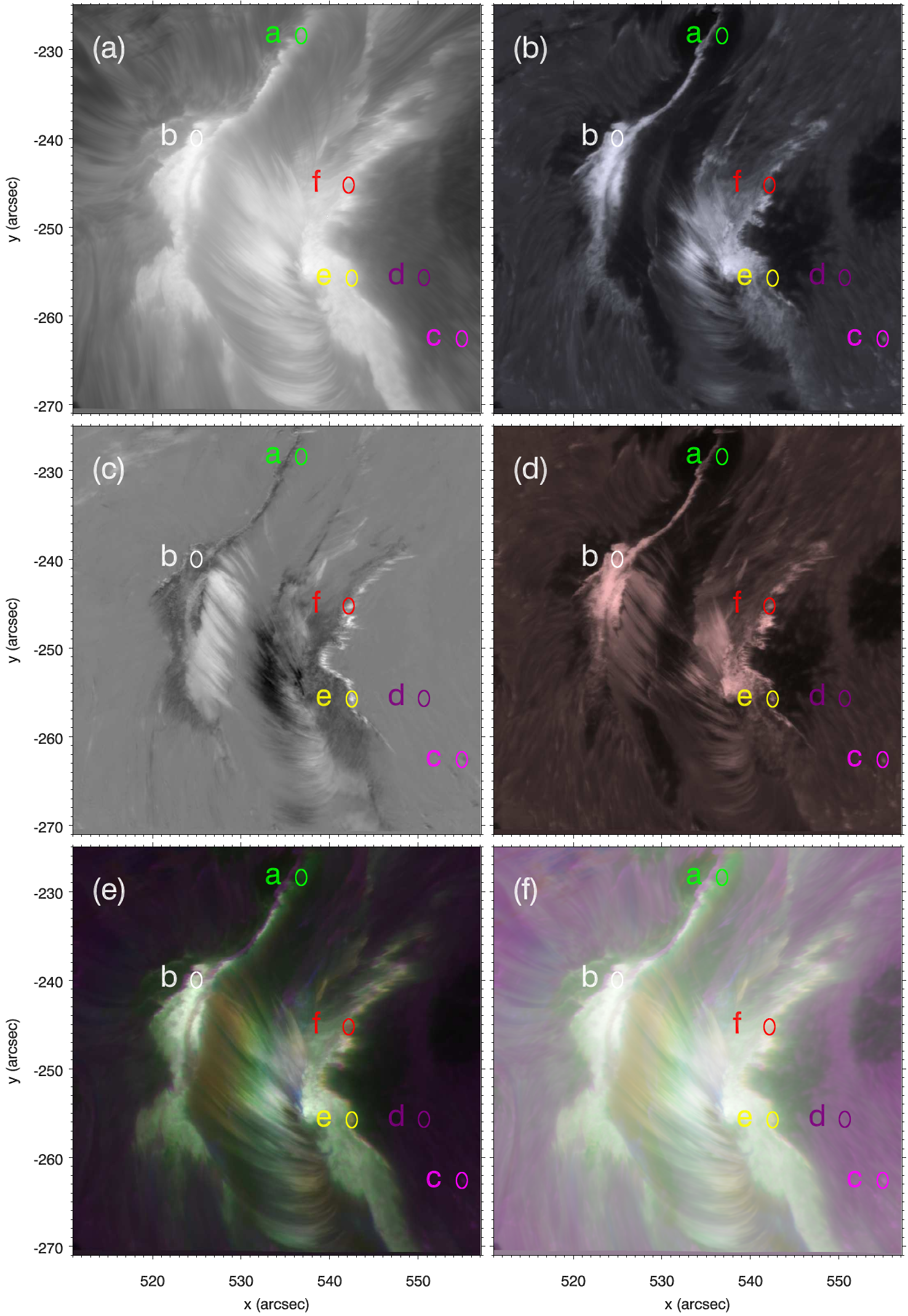}
   \caption{Visualizations of SST/CRISP \Halpha\ observations of an X9.3 class flare that occurred on 6 September 2017. 
   The panels display different visualizations of the flare at 12:11:48\,UT: (a) \Halpha\ line core, (b) \Halpha\ blue  and (d) red wing intensities at $\pm$1\,\AA, (c) \Halpha\ blue-minus-red wing (at $\pm$1\,\AA) difference image, (e) COCOPLOT image taking the full spectral range into account and (f) logarithmic COCOPLOT to display the lower intensity areas. 
   In each panel six locations are marked with circles and the panel letter for which the profiles are displayed in Fig.~\ref{fig:colors}.}
   \label{fig:flarecompare}%
   \end{figure*}

\begin{figure*}
   \centering
   \includegraphics[width=\textwidth]{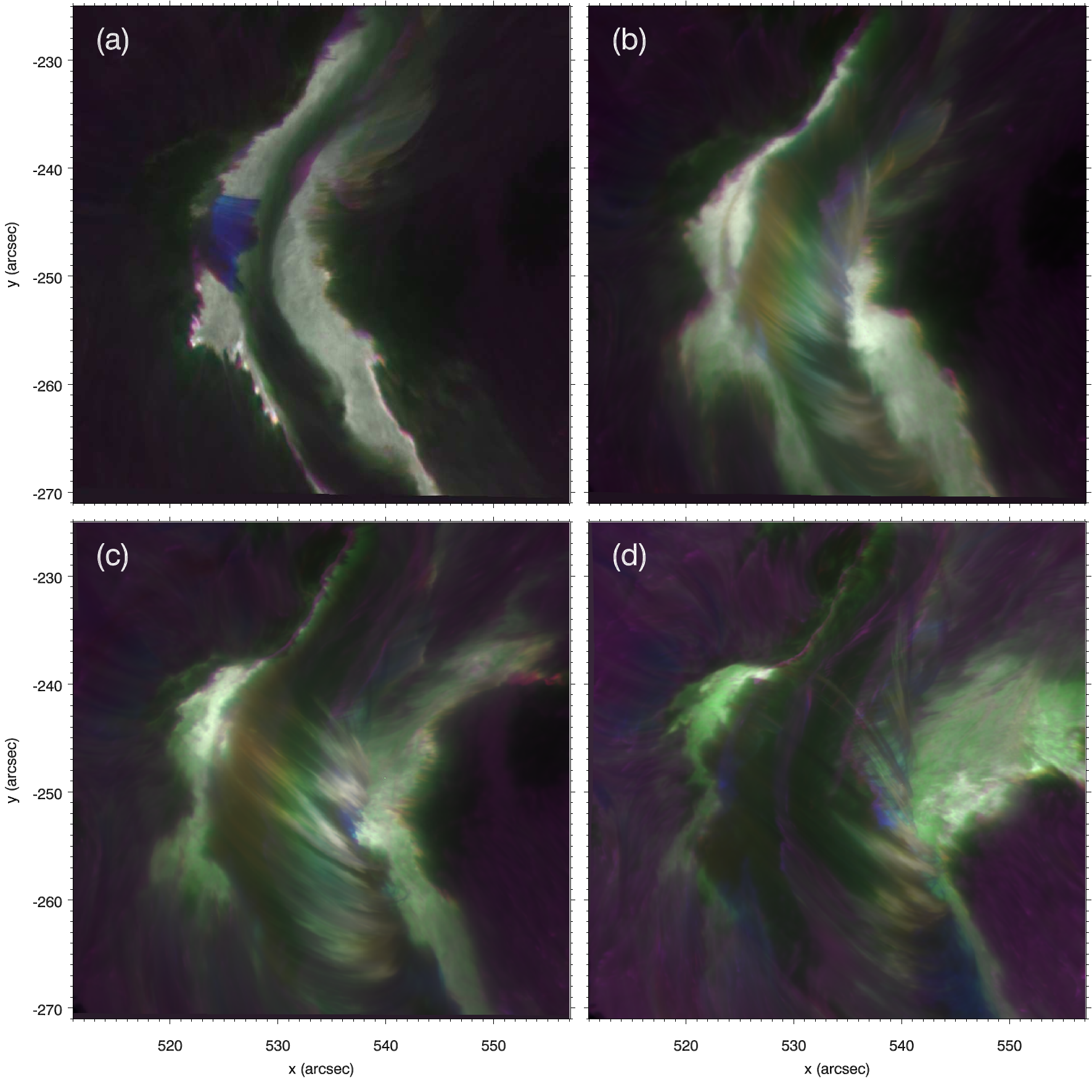}
   \caption{COCOPLOTs of the same flare as in Fig.~\ref{fig:flarecompare}, but at different times during 
   its evolution: (a) 12:00:50\,UT, (b) 12:06:45\,UT, (c) 12:15:03\,UT, (d) 12:31:45\,UT.}
   \label{fig:flareevol}%
   \end{figure*}

COCOPLOTs offer an excellent way of visualizing spectral profile peculiarities that may arise during explosive events like, for instance, the X9.3 class solar flare that occurred on 6 September 2017 in NOAA Active Region AR12673 at \xy{537}{-242} (S09W34).
This flare was observed at the \SST\ \citep[SST,][]{2003Scharmer} with \Halpha\ imaging spectroscopy from the \CRISP\ \citep[CRISP,][]{2006Scharmer}.
The \Halpha\ line was sampled at 13 non-equidistantly spaced spectral positions covering the wavelength range $6563.0\pm1.5$\,\AA\ at a 15\,s cadence. The flare was also observed in \CaIR\ with full Stokes polarimetry. A full account of these data is provided in \cite{2019Quinn}.
Both datasets were reduced again for use in this paper using the SSTRED pipeline as described by \citet{jaime15} and \citet{mats21}, which make use of Multi-Object Multi-Frame Blind Deconvolution \citep[MOMFBD;][]{vanNoort05,mats02}. Additionally, we performed an absolute wavelength and intensity calibration using the solar atlas by \cite{Neckel1984}.

Under quiet (Sun) conditions, the \Halpha\ line exhibits an ordinary absorption profile with average bright wings and a deep line core (e.g.~Fig.~\ref{fig:colors}c and d), while sunspots for instance appear dark throughout the spectral profile.
However, during solar flares, the \Halpha\ line often goes into emission, with line core intensities surpassing those of the line wings and effectively becoming an emission line. In addition, the core of the line can broaden out over more than the limited 3\,\AA~wavelength window of the observations we consider here due to non-thermal broadening, greater formation temperatures, and the Stark effect. 

Figure~\ref{fig:flarecompare} shows the traditional, monochrome images of the flare at particular wavelengths, with the line core shown in panel a and the shorter and the longer wavelengths (i.e. the blue and red wings, respectively) shown in panel b and d, respectively. A red-minus-blue difference image is shown in panel c with a color scale running from dark (brighter in the blue wing) through gray (no difference) to white (brighter in the red wing). 

Let us now demonstrate a use case for COCOPLOT as a quick-look tool with this data. Critically, observers and theoretical solar physicists are interested in chromospheric up-flows and down-flows during solar flares, as those can provide important information about energy delivery throughout the atmosphere, rendering regions of red or blue-shifted emission of particular interest.

To find locations of potential down-flows in the flare, with Doppler-shifted enhancement in the red wings of the line, we may first use quick-look software to examine the \Halpha\ red wing image (Fig.~\ref{fig:flarecompare}d). In this image we can immediately identify the dark sunspots and the average intensity quiet sun, contrasted with nearly all of the flare pixels which appear white in monochrome gray-scale images because they have high values in the red wing of the spectral line.

One might, for example, blink between images in the red and blue wings of the profile (Fig~\ref{fig:flarecompare}b and d) From all of these areas, we are interested in those showing solely enhancement in the red wing, so for the next stage one could create a difference image between the red and blue wing (Fig.~\ref{fig:flarecompare}c).

However, although a red-vs-blue-wing difference image highlights those pixels we are interested in, it also bring out regions that are centrally enhanced, or have normal intensity in one wing and an absorption feature in the other. One must then manually search through the remaining pixels to identify those only enhanced in the red wing, potentially eliminating those that are also centrally enhanced or have absorption in the blue wing.

This quickly becomes a long and tedious process, that can also make it harder to keep in mind the full context of the events in which the profiles are occurring. This situation therefore makes \Halpha\ flare data ideal for use with COCOPLOT images, indeed it was the original motivation for developing the software.

COCOPLOTs tremendously simplify such data analysis.
Figure~\ref{fig:flarecompare}e shows a COCOPLOT image for the same frame, constructed using the filter setup described in section \ref{sec:colors} (see Fig.~\ref{fig:colors}, dashed lines).
Using this COCOPLOT image one can instantly identify the potential regions with down-flows as brighter red.
The relevant pixels are situated predominantly in footpoints of the arcade loops (in the lower part of the image), but also on the outer edge of the bright flare ribbon. Some regions along this ribbon edge are also centrally enhanced, and these appear yellow in the COCOPLOT. A traditional difference image for the red and blue wings would not differentiate between these `red' and `yellow' COCOPLOT profiles. 
Moreover, additional context and features are immediately conveyed via the COCOPLOT that may not be evident from other methods. Let us look further at some of these features of the flare, focusing in particular on the comparison between our COCOPLOT results and other techniques such as difference images or a set of monochrome figures. 

\begin{figure*}
   \centering
   \includegraphics[width=\textwidth]{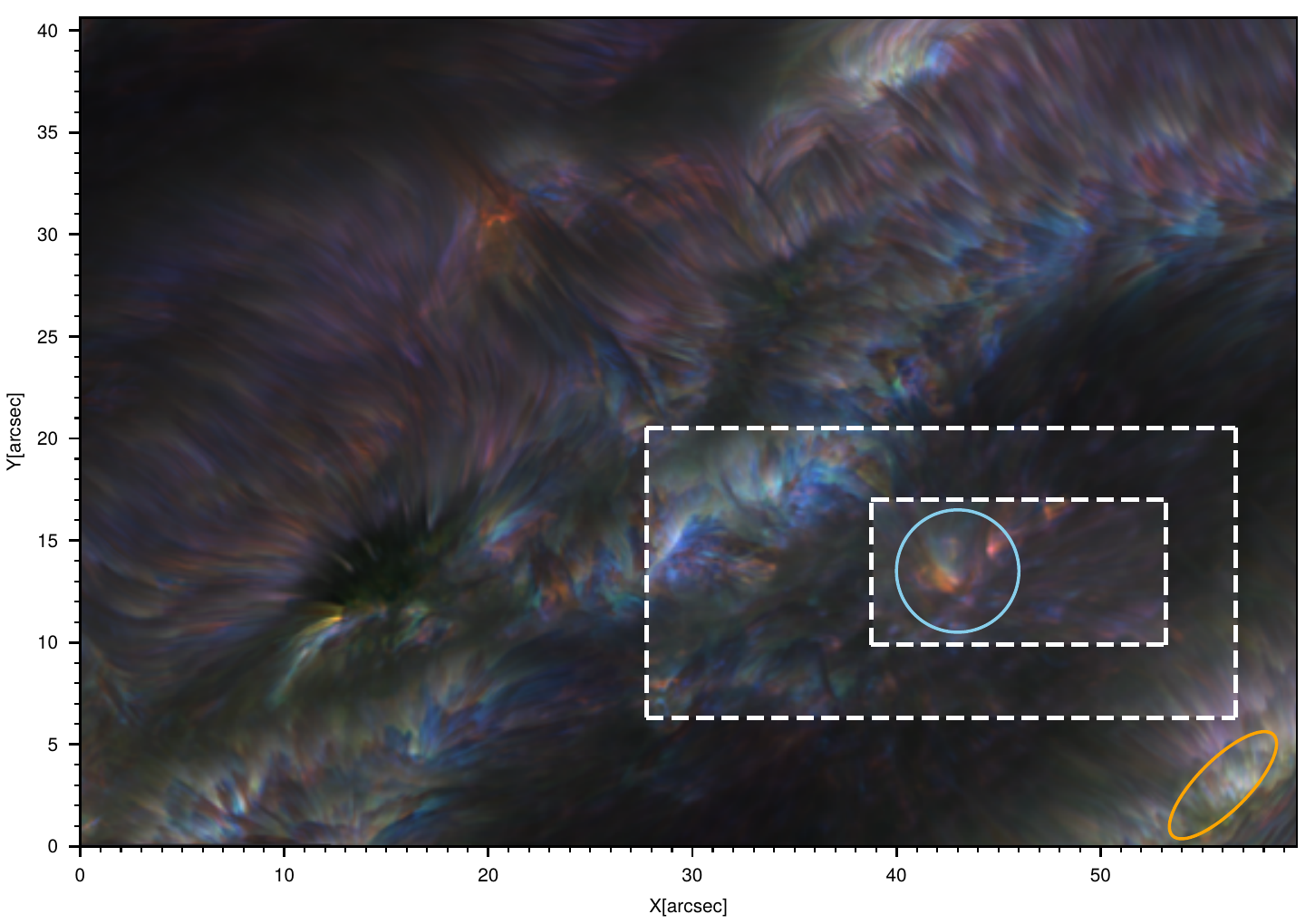}
   \caption{COCOPLOT image of the data presented in \citet{2019Robustini} displaying a supergranular structure observed in the nominal line center of \CaIIK. The main image and the two insets (dashed white rectangles) have different intensity thresholds to enhance the color contrast. The smaller rectangle marks the cutout from Fig.~\ref{fig:clustering}, top panel, while the colored circle and ellipse highlight features discussed in the main text.}
   \label{fig:cocoCaK}
\end{figure*}

The flare began before the observational sequence and the seeing quality of the observations was not consistent throughout the sequence, however there are a number of excellent images in the time-series.
Figure~\ref{fig:flareevol} presents four (taken at (a) 12:00:49\,UT, (b) 12:06:44\,UT, (c) 12:15:02\,UT, and (d) 12:31:45\,UT) to provide an overview of the uses of COCOPLOTs. 
%
From these frames one can see that this flare evolves in some typical ways:
\begin{itemize}
    \item The main bodies of the flare ribbons feature bright, broad \Halpha\ line profiles due to collisional broadening, high ionization increasing the free electron densities and Stark effect, as well as high temperatures at these times. Therefore, the ribbons appear white during the earlier parts of the flare (Fig.~\ref{fig:flareevol}a and Fig.~\ref{fig:flareevol}b). As time progresses and the ionization degree and temperatures return to pre-flare levels, the \Halpha\ profiles in the flare ribbons become narrower, but remain centrally enhanced therefore appearing green at later times (see Fig.~\ref{fig:flareevol}c and Fig.~\ref{fig:flareevol}d). This behaviour would be somewhat intuitive to the experienced researcher in solar flares, but is naturally displayed and lain bare for all to see using a COCOPLOT, even when focusing on other behaviour.
    \item There are two bright flare ribbons that appear in the early stages for the flare. These ribbons start out close together, but expand and move apart as time progresses. A large amount of the expansion of the flare ribbons has already occurred by the time shown in  Fig.~\ref{fig:flareevol}a. The separation of the western and eastern flare ribbons increases from Fig.~\ref{fig:flareevol}a, to b, to c. For example, the eastern ribbon expanded into an area around \xy{540}{$-$260} during the times between the images shown in panel a and panel b, and into the area around (545\arcsec,$-$238\arcsec) in the time between panel b to panel d. The flare ribbons begin to shrink at the end of the flare (compare the western flare ribbon around (520\arcsec,$-$255\arcsec) in Figs.~\ref{fig:flareevol}c and d.
    \item A post-flare loop arcade forms between the two ribbons after some time (the arcade is not present in Fig.~\ref{fig:flareevol}a, but emerges in panel b between the two ribbons, for example around (528\arcsec, $-$250\arcsec)). The loops of this arcade later expand to appear open at time where the temperatures in these loops becomes too hot to contain neutral hydrogen (see the expansion of the arcade identified in Fig.~\ref{fig:flareevol}b, above, which has expanded by the time at which panel c was taken as the ribbons separated, this ribbon continues to expand and is seen to begin fading by Fig.~\ref{fig:flareevol}d). 
\end{itemize}

Some more exotic behaviors in this flare become immediately evident when using COCOPLOTs, which are not so clear from looking at difference images or scrolling through the spectral dimension:
\begin{itemize}
    \item In the western part of the sunspot group, a light bridge harbors the base of some recurrent fan-shaped jets (see Fig.~\ref{fig:flareevol}a, at \xy{520}{$-$225} with the light bridge evident in the other panels at the same location). The bulk of these jets usually consists of cold plasma ejected from the lower chromosphere \citep{2018A&A...609A..14R} and its appearance occurs before the flare. When the flare ribbon expands, the jets are suppressed and thus no longer visible in the COCOPLOTs and intensity images. The light bridge is maintained beneath the structure and can be identified in the COCOPLOTs as a dark patch or line, within the bright white ribbon (panels b and c). Towards the end of the time series there is some evidence of the jets restarting (or re-appearing), with faint blue wing emission being seen in panel d at the same location, once the ribbon has moved away. In the COCOPLOT image the light bridge at the base appears purple (panels a and b), indicating absorption profiles with enhancement in the wings. The fan shaped jet collapses after the flare ribbon meets its base. The profiles in the collapsing fan-shaped jet are characterized by strong broad back-lighting from the flare ribbon and a red-shifted line core in absorption due to the material in the fan jet that is in front of the flare in the line of sight. The absorption extends into the red wing nearly up to 1.5\,\AA\ from line center. As a result the red and green channels have lower values, leaving the area to appear a bright blue in the COCOPLOT. This COCOPLOT image was the original inspiration that lead to the depth study of the feature presented in \citet{2022Pietrow}, which produced the first observational estimate of the mass contained in such a structure.
    \item The leading edges of the flare ribbons are seen to produce mainly red-shifted emission while they move apart. These regions appear reddish in the COCOPLOTs, or yellow/orange where the profiles are also centrally enhanced (e.g. the west side of the western ribbon in Figs.\ref{fig:flareevol}a around \xy{540}{$-$255} and (535\arcsec,$-$245\arcsec), \ref{fig:flareevol}b around (543\arcsec,$-$260\arcsec) (538\arcsec,$-$254\arcsec) and the north-eastern part of the eastern ribbon in Fig.~\ref{fig:flareevol}c around (553\arcsec,$-$240\arcsec)). However, one should be very careful to avoid inferring down-flows in the flare ribbons from red-shifted profiles that result from projection effects due to viewing angle (considering $\theta=53^{\circ}$ ($\mu=0.6$)) and the varying height of the chromosphere (see the left part of the lower flare ribbon in Fig.~\ref{fig:flareevol}a, Fig.~\ref{fig:flareevol}b, and \ref{fig:flareevol}c). Using this projection effect on the northern part of the eastern ribbon (around \xy{536}{$-$228} in panel b and c) we can see that the bright flare ribbon emission (white) appears to be overlapped by the quieter chromospheric \Halpha\ emission (purple) which obscures part of the flare ribbon and therefore, given the viewing angle, can be deduced to be coming from a higher point in the atmosphere. Less broad, centrally enhanced emission (green wisps around (538\arcsec, $-$228\arcsec) are coming from the area above where the flare ribbon has formed. This picture of the formation heights of the flare ribbon agrees with earlier reports suggesting lower formation heights of the flare ribbon in emissions such as \Halpha\ \citep{2017Druett} and white light \citep{2012MartinezOliveros, 2018Druett}. This could only be described with a large number of words or the use of multiple images via traditional methods, but can be instantly gleaned from a COCOPLOT.
    \item Spicule-like jets poke through the flare ribbons throughout the early stages of the flare, seen as a purple-on-white `polka-dot' pattern in unsaturated pixels throughout the main bodies of both flare ribbons that appear white in the COCOPLOT Fig.\ref{fig:flareevol}a. When the emission is less intense and broad at later times (darker and greener), we are able to see these features more clearly, giving the purple on green appearance of a `lavender-field' flare ribbon (panels c and d). Although there is some imprint of these as darker or lighter features in the monochrome images (compare the purple features on the flare ribbon in Fig.~\ref{fig:flarecompare}d with the speckling in the same flare ribbon in Fig.~\ref{fig:flarecompare}a), they are much more clear using a COCOPLOT. In fact we identified these features as a result of using COCOPLOTs. An analysis of their properties is being performed and will be presented in a separate paper.
\end{itemize}
These examples show that COCOPLOT is a versatile method that can be used to identify and expose particular spectral behaviors, highlighting various (evolutionary) features in the process.

\subsubsection{COCOPLOTs versus $k$-means profile clustering} \label{sec:CaK}

\begin{figure}
   \centering
   \includegraphics[scale=1]{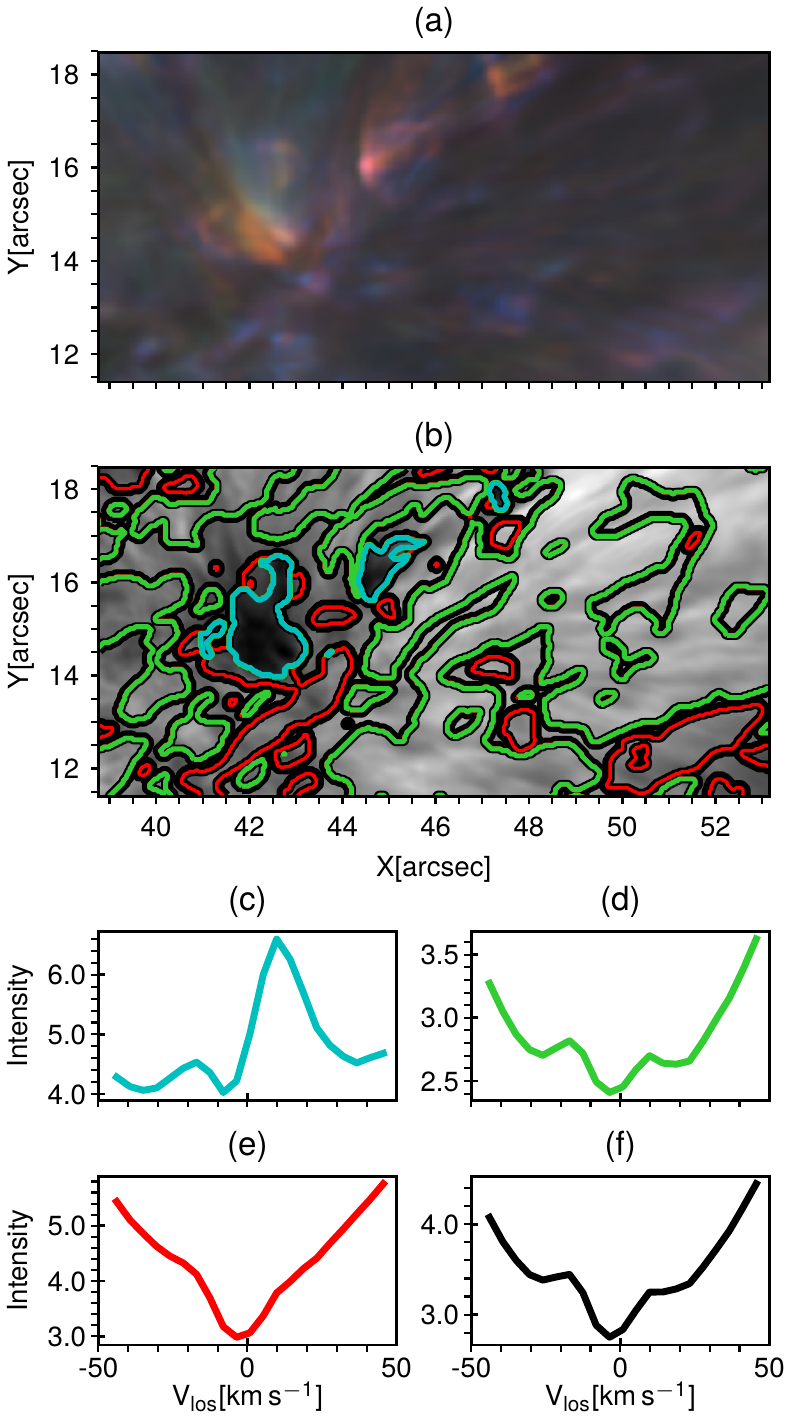}
   \caption{COCOPLOT and $k$-means clustering comparison of the smaller sub-field-of-view in Fig.~\ref{fig:cocoCaK}. Panels (b)--(f) are taken from \citet{2019Robustini} : (a) COCOPLOT, (b) inverted \CaII\, \Kthree\ image with colored contours marking the spatial distribution of profiles belonging to a particular cluster, (c)--(f) cluster-average profiles with identical color coding as in panel b.}
      \label{fig:clustering}
\end{figure}

Another way to provide a spatial overview of spectral information is by 2D maps obtained from spectral profile classification. Some recent applications in solar physics include \citet{2017A&A...602A..80D}, \citet{2017ApJ...835..156N} and \citet{2019A&A...621A..35L} using $k$-nearest neighbour, and  \citet{2011A&A...530A..14V}, \citet{2018ApJ...861...62P} , \citet{2019Robustini}, and \citet{2021Woods} using $k$-means clustering. We will focus on the latter and use some data previously analyzed with the clustering technique by one of the authors for a comparison with COCOPLOTs. 

\cite{2019Robustini} present chromospheric observations in the narrow band of \CaIIHK, targeting a supergranular structure with a radial arrangement of fibrils and located next to a small active region. The data were obtained at the SST on 20 April 2017 with the CHROMospheric Imaging Spectrometer \citep[CHROMIS,][]{2017psio.confE..85S} at a 15\,s cadence. Figure~\ref{fig:cocoCaK} displays a COCOPLOT of the full field-of-view of these observations. The authors applied $k$-means clustering to the \CaIIK\ spectral profiles belonging to the portion of quiet Sun shown in the smaller inset of Fig.~\ref{fig:cocoCaK}, and Fig.~\ref{fig:clustering}b shows the resulting label distribution on the inverted line core intensity map. Here, the chosen number of cluster centers is four and for each center the corresponding profile is shown in panels c--f.

These profiles exhibit the typical features characterizing the regions highlighted with the same label (color). In this type of classification the user does not prescribe a characteristic profile, but rather the number $k$ of clusters (yielding one cluster-average profile for each). Therefore the same number of cluster centers may produce different cluster-average profiles depending on the selected field-of-view. The higher the number of clusters, the closer a profile in the cluster will be to the cluster-average profile. 
However, the $k$-means algorithm for context images as Fig.~\ref{fig:clustering}b is meant to expose the general behaviour of the region of interest and not the detailed spectral behavior. Moreover, this algorithm performs better with a limited number of cluster centers. 
The wavelength range of the data cube used for the clustering in \cite{2019Robustini} is $3933.68\pm 0.58$\AA, corresponding to the deeper part of the \CaIIK\ spectral line. Although pure absorption profiles, as in \Halpha, can be observed (see Fig.~\ref{fig:clustering}e), in quiet Sun this spectral line usually has a complex shape (see Fig.~\ref{fig:clustering}d and f) which makes clustering techniques particularly appealing. Following the classification in \citet{1904ApJ....19...41H}, \CaIIK\ exhibits two emission peaks, one in the red (\KtwoR) and one in the blue wing (\KtwoV), and a core in absorption (\Kthree).

We produced color collapsed images out of the same data used in \cite{2019Robustini} by applying two narrow normal blue and red filters ($\sigma=0.006$\AA) centered on the wavelength corresponding to the \KtwoR\ and \KtwoV\ features and a broader green filter ($\sigma=0.1$\AA) at the \Kthree\ feature to identify regions with central enhancements that may or may not be slightly Doppler-shifted. The result is shown in Fig.~\ref{fig:cocoCaK}.
%
The two insets shown in this figure have different upper and lower intensity thresholds in order to bring out the color contrast in the COCOPLOTs. The field-of-view of the smaller inset is the same as that shown in Fig.~\ref{fig:clustering}a--b.

We can easily spot the regions where the line core is most enhanced (green prevalence) as highlighted by the orange ellipse or where a strong peak asymmetry is present (red or blue dominates). 
Figures 9 and 10 of \cite{2019Robustini} show that the chromospheric brightening observed in \CaIIK\ corresponds to a region of red-over-blue \Ktwo\ peak asymmetry and blue-shift of the nominal line center, as highlighted by the azure contours in Fig.~\ref{fig:clustering}c. COCOPLOTs show that this region (cf.~azure circle in Fig.~\ref{fig:cocoCaK}) is characterized by a red-orange color, corresponding to a \KtwoR\ enhancement, but also by a greener and bright purple component. Since each pixel in a COCOPLOT is treated separately, it can reveal a complex behaviour of the profiles at a higher spatial variation than could be obtained with $k$-mean clustering, even when increasing the number of clusters.

\subsubsection{Feature identification in IRIS data} \label{sec:IRIS}
\begin{figure}
   \centering
   \includegraphics[width=\columnwidth]{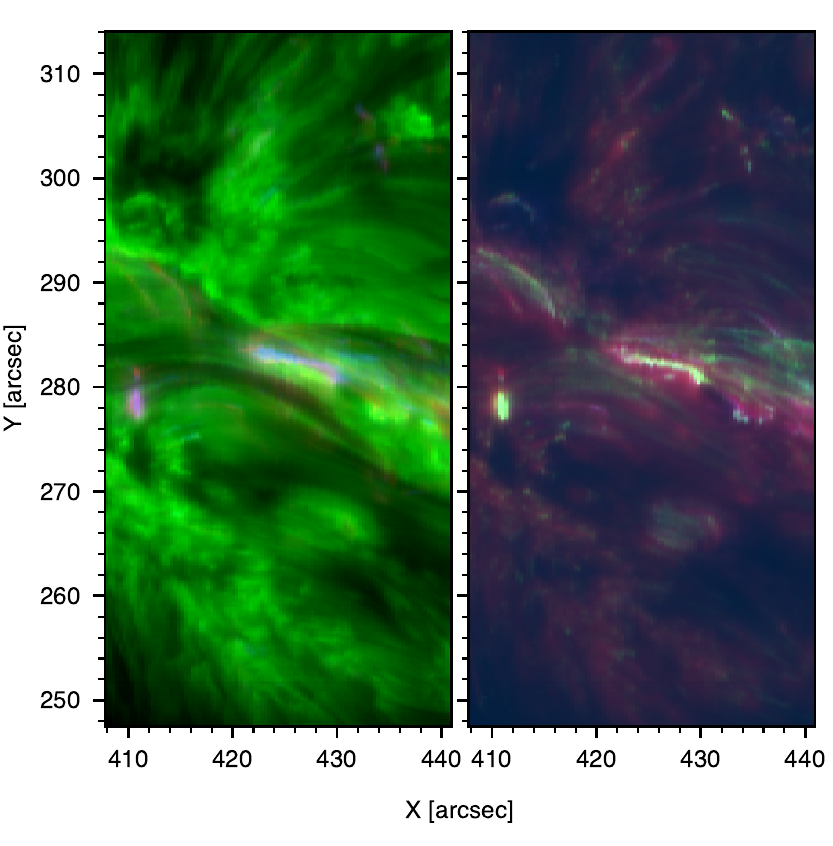}
   \caption{COCOPLOT of an IRIS spectrograph raster highlighting feature visibility in the \MgIIk\ line ({\it left\/}) and in \CII, \SiIV\ and \MgII\ \ktwoR\ ({\it right\/}).
   }
   \label{fig:iris_sg}%
\end{figure}

COCOPLOTs are ideal for spectral profile classification and they can, by extension, also be used for solar feature identification.
Figure~\ref{fig:iris_sg} provides an example based on a 92-step IRIS raster of NOAA AR 12089 observed on 15 June 2014.
This active region has previously been analyzed in 
\citet{2015ApJ...812...11V} 
to investigate the correspondence between \EBs\
\citep[][strong enhancements in the \Halpha\ wings]{%
1917ApJ....46..298E} 
and \UVBs\ \citep[][broad enhancements of the \SiIV\ lines]{%
2014Sci...346C.315P, 
2018SSRv..214..120Y}. 
In addition to strongly enhanced and broadened \SiIV\ lines, \UVBs\ are also characterized by enhanced \CII\ lines around 1335\,\AA\ and an enhancement of the \MgIIhk\ wings, while the \MgIIhk\ cores remain relatively undisturbed.
With an appropriate choice of filters COCOPLOT can therefore uncover UVBs in IRIS observations, as evidenced by both panels in Fig.~\ref{fig:iris_sg} (displaying the same field-of-view at the same time).
The left-hand panel highlights the \MgIIk\ line, with filters positioned at $\pm$0.5\,\AA\ from the \MgIIk\ core (blue and red, respectively), as well as at \kthree\ (green). 
The chosen filter widths are sufficiently narrow ($\sigma$ = 0.1\,\AA) and their positioning such that the \MgII\,\kthree\ core contributes minimally to 
the blue and red filters, while the typical \ktwo\ wing enhancements from UVBs have minimal impact in the green filter.
The right-hand panel picks out the red wing of \CII\,1335.7\,\AA\ (blue), \SiIV\,1402.8\,\AA\ (green) and \MgII\,\ktwoR\ (red) with somewhat broader filters ($\sigma$ = 0.25\,\AA).

Several features become evident in these visualizations. 
\UVBs---the most prominent of which around \xy{412}{278}---light up in purple indicating visibility in the \ktwo\ peaks, but not in the \MgIIk\ core, as one would expect 
\citep[cf.~e.g.][]{%
2015ApJ...812...11V, 
2016ApJ...824...96T, 
2016A&A...593A..32G, 
2017A&A...598A..33L}. 
The same event is colored green-yellow in the right-hand panel, indicating visibility also in \SiIV\ and to lesser extent in \CII.
Overlying canopy fibrils are dark-green to black in the left-hand panel, suggesting they are dark features in the \MgIIk\ core and near-invisible in the wings of the line, while the plage stands out in green as predominantly bright in the line core with negligible contribution in the wings.
Partly coinciding with the dark canopy fibrils the right-hand panel exhibits green to cyan filamentary structures (some of which are seemingly emanating from the UV bursts) indicating these are bright structures \SiIV\ and again to a lesser extent in \CII, while invisible in the red \MgIIk\ wing.

\subsection{Visualizing temporal evolution} \label{sec:time}
In the previous sections we have shown that the COCOPLOT principles work intuitively by giving a summary of the available spectral data, however, with an appropriate choice of filters COCOPLOT can provide a visualization of essentially any 3D data cube.
Here we discuss visualization of time series as an example, being the most obvious case after imaging spectropolarimetry (either applied to individual Stokes parameters, or some useful combination of different Stokes parameters, such as their ratios), but this is by no means the only alternative use of this method.
\begin{figure*}
   \centering
   \includegraphics[width=\textwidth]{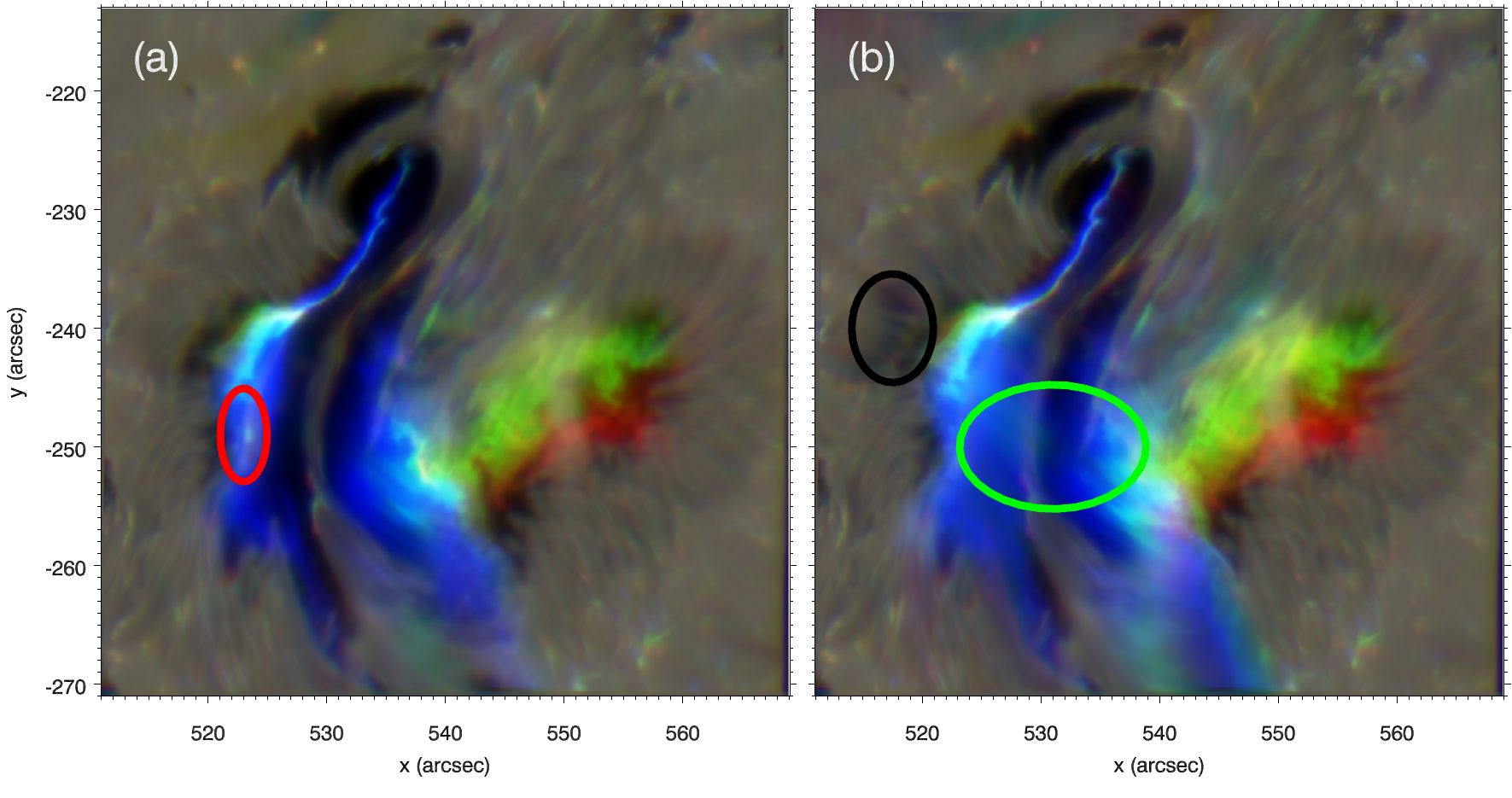}\label{fig:timer}
   \caption{Temporal COCOPLOTs of the same X9.3 class flare as in Figs.~\ref{fig:flarecompare} and \ref{fig:flareevol}: (a) \Halpha$-$1.0\,\AA\ blue wing, (b) \Halpha+1.0\,\AA\ red wing.
   Areas showing enhancement early in the time series are encoded in blue, those around the middle in green and those towards the end in red.
   Features of interest are highlighted with colored ellipses and discussed further in the main text.}
   \label{fig:flaretime}%
\end{figure*}

Time series analysis is a common element of solar physics research, e.g.\citet{2012Viall}.
In particular in active regions, where considerable changes may take place over the course of a few frames, while larger-scale evolution proceeds more gradually, COCOPLOTs present an intuitive quick-look summary of the location and timing of various changes.
By applying for instance three Gaussian filters (covering the entire time series) to a time series of monochromatic images, or the net emission in a set of spectral positions ($n_x,n_y,n_t$) areas that show high signals early (blue), in the middle (green) and late (red) in the time series (and combinations of these) can be differentiated at a single glance. On the other hand, continuous processes will color white, while regions with strong signals early and late in the time series, but not in the middle will show up as purple. 
Moreover, these images provide summaries of time series in instances where a video or many individual frames may not be feasible, for example in a poster or a print-out of a paper. It is essential to realize that the colors are not single points on a time-line (i.e.~there is no `time-dependent color bar'), but rather the color encodes the range of times at which the signal was high, and the figure is a map of this. 

Figure~\ref{fig:flaretime} presents temporal COCOPLOTs of the same X-class flare as discussed in section~\ref{sec:Halpha}, with the RGB image for the wavelength points $\mp$1.0\,\AA\ from the line core in the left- and right-hand panels, respectively. Comparing these two images also provides a summary of the similarities and differences in the \Halpha\ wing behavior during the flare. For example, both wings show the same general expansion and separation of the eastern and western flare ribbons, seen in the pattern of blue to green to red coloration, over time. The flare arcade loops reaching down to the eastern ribbon in the early to middle of the observational sequence (blue-cyan color, highlighted by the green ellipse in Fig.~\ref{fig:flaretime}b) show emission enhancement only in the red wing of the line. Inspecting the detailed line profiles reveals that these indeed correspond to down-flowing (i.e. Doppler red-shifted) bright material during the early-to-middle section of the time sequence. 
A dark patch in the far east of the field-of-view is evident in the red wing COCOPLOT (area inside the black ellipse in panel b), but not the blue wing. Upon inspecting the profiles in this region, it appears there is also a downflow of red-shifted material coming in from the east of the field-of-view, however, in absorption and persisting throughout the whole observation. In the blue wing image (panel a), one can more clearly see the restarting of the recurrent fan jet (red ellipse) that was discussed in section \ref{sec:Halpha}. This occurs late in the sequence above a light bridge in the western ribbon and it therefore shows a reddish contrast compared with its surroundings. These are only a few of the many features that illustrate the use of temporal COCOPLOTs and other features could be identified or highlighted by changing the filters to focus on more specific times in the sequence.

\begin{figure*}
   \centering
   \includegraphics[width=\textwidth]{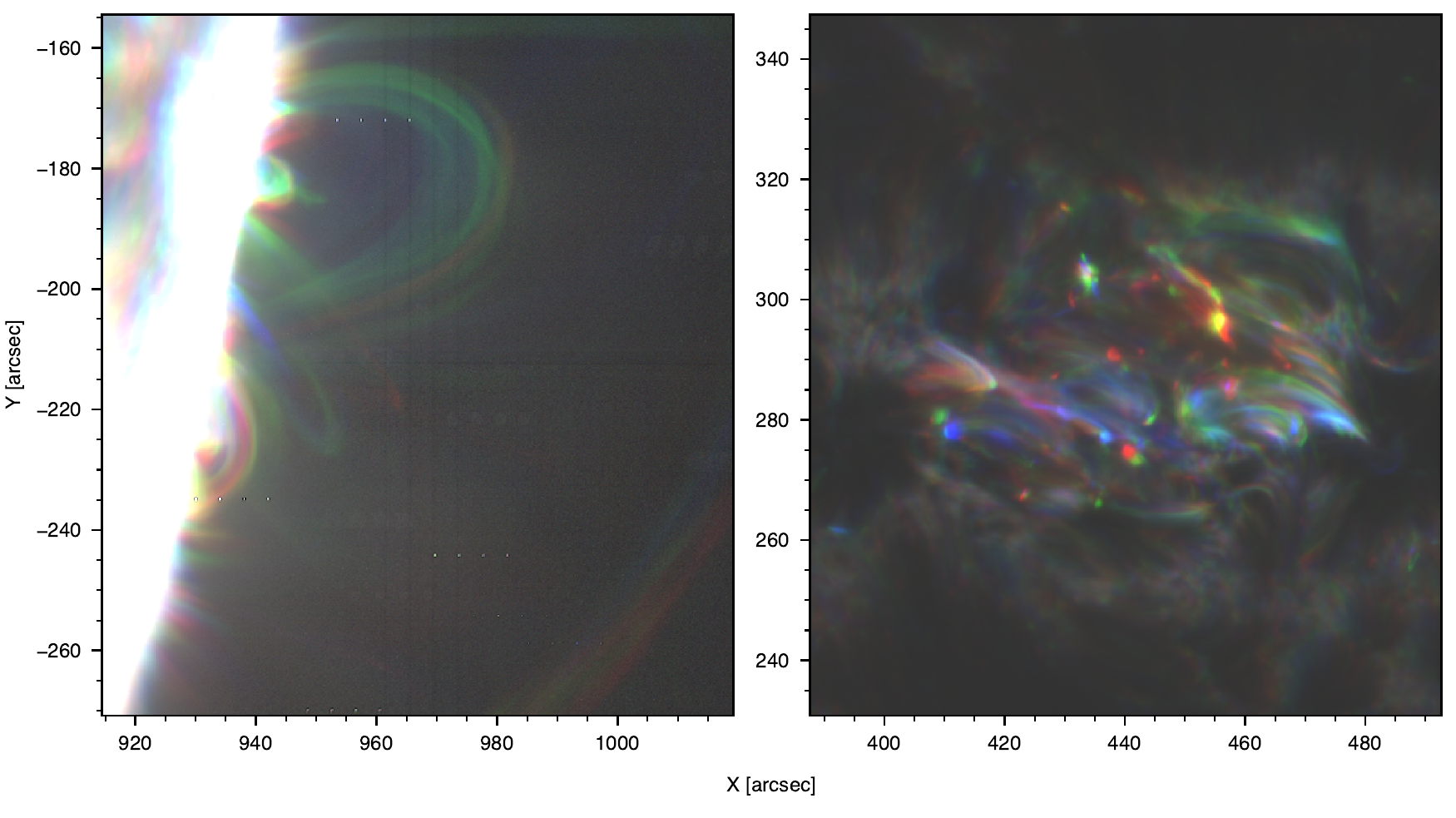}
   \vspace{-4ex}
   \caption{COCOPLOTs of IRIS slit-jaw image time series in \CII\ ({\it left\/}) and \SiIV\ ({\it right\/}) using band filters splitting the observations in three equal parts (blue, green and red filter for the first, middle and last third, respectively).
   }
   \label{fig:iris_sji}%
\end{figure*}

Figure~\ref{fig:iris_sji} showcases another application of temporal COCOPLOTs, based on IRIS \CII\ slit-jaw images of NOAA AR 12035 observed on 24 April 2014 at the west limb (left-hand panel) and \SiIV\ images of AR 12089 observed on 15 June of the same year (right-hand panel).
Both images were constructed using box-car `band' filters of, respectively, 90 and 63 minutes wide, splitting each time series in three equal parts.
The left-hand panel shows the use of COCOPLOT in visualizing the evolution of off-limb structures such as spicules (around \xy{930}{$-$250}), low-lying loops (around \xy{940}{$-$225}) and coronal rain and loops around \xy{980}{$-$180}. 
The bright white band at the limb is an inevitable result from intensity clipping to bring out off-limb features, however, solar rotation effects are still visible inside the limb around, e.g. \xy{920}{$-$160} where a similarly shaped bright patch is predominantly blue at (915\arcsec,$-$160\arcsec), green at (918\arcsec,$-$160\arcsec) and red at (921\arcsec,$-$160\arcsec).
The larger coronal loop-like structures in the upper middle of the field-of-view (around \xy{980}{$-$180}) outline both persistent loops and traces of coronal rain.
As the smaller loop structures are predominantly blue, with larger structures transitioning through green to red, this COCOPLOT color coding shows a combination of increasingly higher-located coronal loops lighting up and coronal rain falling down from higher heights as the time sequence progresses.
Several coronal rain traces undergoing similar evolution are also faintly visible in the lower-right corner (around \xy{1005}{$-$250}).
In both cases the individual coronal rain loops are not immediately distinguishable from coronal loops without rain, however the temporal COCOPLOT was used as a quick-look tool to identify evolving and expanding loop systems that are known to be likely locations of coronal rain, which was indeed confirmed by further investigation.

The right-hand panel exhibits a large number of compact brightenings in the \SiIV\ slit-jaw channel that are likely UV bursts and that have one of the primary colors.
This indicates that most do not persist throughout the time series, while others evidently last longer (e.g. the yellow feature around \xy{455}{297}).
One can also see at a single glance that activity is primarily centered around \xy{420}{280} in the early part of this time series (cf.~the blue and green compact brightenings and blue filamentary loops), while in the second half of the time series activity increases in the upper right part of the active region, as indicated by the multiple red compact brightenings around \xy{450}{290}, as well as green and yellow filamentary loops around \xy{465}{285} and (465\arcsec,313\arcsec).

\section{Conclusions} \label{sec:conclusions}

We have presented the COlor COllapsed PLOTting quick-look, figure, and context image software. 
COCOPLOT enables conveying information from 3D data cubes via 2D color images and is a versatile method that can be used to identify and expose particular spectral and temporal behavior, as well as displaying and highlighting solar features and phenomena as desired.
We have presented a variety of use cases where this software can be an excellent tool as a spectral line quick-look method, helping the user to identify regions of interest more rapidly, as well as instantly displaying other interesting behaviors that may not have been considered initially.
COCOPLOTs are also outstanding context images that can provide a fuller sense than monochrome images would of the environment in which detailed analysis is subsequently performed.
However, since COCOPLOT colors do not correspond one-to-one to single spectral positions or time values, we stress the need to sufficiently explain the meaning of these colors to ensure that the information encoded therein is interpreted correctly. 

An obvious limitation of the method is that it allows visualizing only three regions of critical variation in the spectral profile or temporal evolution and in certain contexts other methods (such as $k$-means clustering or analysis of monochrome images at particular wavelengths) may be more appropriate.
Yet, while they are no substitute for detailed analysis of the full spectral and temporal domain of a data set, COCOPLOTs offer an effective way of visualizing information contained in higher data set dimensions and can thereby guide analysis to regions of interest that might otherwise have been overlooked.

A 2-filter COCOPLOT will also convey more information than a monochrome or color-scaled difference image via the intensities of the two color filters used, and can be achieved by setting all of the third filter's values to zero. This could be useful, in particular, for any colorblind users, however in this first generation of the code no formal colorblindness corrections are explicitly included. We have also not yet included any consideration of an alternative band set such as CMY, or a 4-filter setup such as those used for the separate bands in PM-NAFE.

We have made COCOPLOT publicly available online
for both IDL and Python\footnote{\url{https://github.com/mdruett/COCOPLOT} or through PyPI 'pip install cocoplots'}. 
The software has been written such that it can be used as stand-alone package, while also facilitating integration in other quick-look software (indeed, an interface to call COCOPLOT is currently being implemented in CRISPEX).
Several publications are already planned that make use of this software, preliminary results of which are briefly presented in the results section of this method paper (Sec.~\ref{sec:results}). 
It is our hope that COCOPLOT will be of benefit to the community and allow to quickly access the desired information in complex data sets, offering in addition a means to present those multi-dimensional data more concisely.

\section{Data Availability}
The SST data underlying this article will be shared on reasonable request to the corresponding author. The IRIS data is available via the Heliophysics Knowledge base  (\url{https://www.lmsal.com/heksearch/}).

\section*{Acknowledgements}
MD is supported by FWO project G0B4521N, and The Swedish Research Council (grant number 2017-04099).
AP \& FC are supported through the CHROMATIC project (2016.0019) of the Knut and Alice Wallenberg foundation
GV is supported by a grant from the Swedish Civil Contingencies Agency (MSB).
CR is supported by the European Union’s Horizon 2020 research and innovation programme under grant agreement No. 759548 (SUNMAG).
The \SST\ is operated on the island of La Palma by the Institute for Solar Physics of Stockholm University in the Spanish Observatorio del Roque de los Muchachos of the Instituto de Astrof\'{i}sica de Canarias. 
The Institute for Solar Physics is supported by a grant for research infrastructures of national importance from the Swedish Research Council (registration number 2017-00625). 
IRIS is a NASA small explorer mission developed and operated by LMSAL with mission operations executed at NASA Ames Research center and major contributions to downlink communications funded by ESA and the Norwegian Space Centre.
We are grateful to Sara Esteban Pozuelo for participating in the SST observations and to Mark Cheung and Charles Kankelborg as IRIS planners.
During our investigation we extensively used CRISPEX \citep{2012VissersCrispex, 2018arXiv180403030L}, SOAImage DS9 \citep{2003DS9} and the CRISpy library \citep{pietrow19} (Now part of the ISPy library \citep{ISPy2021}). 
This work benefited greatly from the feedback from presentations and discussions held with various members of the Institute for Solar Physics at Stockholm University. The concept also owes a lot to the creation of a Bachelors project on solar flares by Malcolm Druett for Veronika Borgstr\"{o}m, a student at Stockholm University, and the subsequent discussions with her regarding her project.

We thank the anonymous reviewers for their input, from which the paper has benefited greatly.


\bibliographystyle{mnras}
\bibliography{cocoplot}

\appendix

\newpage

\section{COCOPLOT procedures and functions} \label{app:functions}
The COCOPLOT code is available from the \href{https://github.com/mdruett/COCOPLOT}{project github}\footnote{This Python package can be installed by running 'pip install cocoplots'.}). 
The code contains the following functions and procedures, and each routine in the code has a file header describing its full functionality.

\subsection{Creating RGB filters}
In order to create an RGB COCOPLOT image, a filter matrix must be supplied to collapse the data with.
The {\bf cocofilter} function can create such filter matrix and takes as argument a set of spectral position values as a 1D array of size $n_{\lambda}$. If the user does not supply these position values, then equidistant spacing along the spectral axis is assumed, using set of integers from 0 to $n_{\lambda}-1$. The user can also specify the filter type and the position and width of the R, G, and B channel filters as described below. With this release, we provide three types of filter in {\bf cocofilter}, `single', `band', or `normal':
\begin{itemize}
\item The `single' filter is called along with three single values for the individual spectral positions desired, each filter has a weighting of 1 at this spectral position and zero at all others. 
\item The `band' filter requires supplied three two-element arrays denoting the start and end subscripts for the range of spectral positions desired. Each filter has a weighting of $1 \div$ band width at each of its spectral positions and is zero at all other positions.  
\item The `normal' filter requires three two-element arrays with the mean and standard deviation for each filter. Each filter weighting is then given by the value of the normal distribution probability density function at this spectral position (cf.~Fig.~\ref{fig:filter}). If positions are not supplied then the default values are set with means equal to the lowest, central, and highest values of the spectral range, with a standard deviation such that $2\times1.96$ standard deviations covers the entire spectral range. 
\end{itemize}
To view the effects of filters on spectral profiles, one can use the {\bf cocofiltplot} function, which takes the spectral profile, the filters, and optionally the wavelengths of the spectral positions. The panels of Fig.~\ref{fig:colors} were generated using this function. 

\subsection{Visualizing COCOPLOT images}
The primary function for visualizing COCOPLOT images is {\bf cocoplot}, which takes the 3D data cube of size [$n_x, n_y, n_{\lambda}$] and filters of size [$n_{\lambda}, 3$], e.g.~the output from {\bf cocofilter}, as required inputs. 
In turn, {\bf cocoplot} calls the function {\bf cocorgb} which actually performs the color collapsing.
Within this function the data cube values are multiplied by the filter values at each spectral position. Then, for the results of the multiplications of each filter's values with the data cube values, a total is taken over the spectral dimension. Thus the operation is akin to taking the vector dot product of the filter values and the spectral profile values at a given spatial point. This computation is performed at each pixel for each filter, generating an array of size [$n_x, n_y, 3$] containing real numbers. 
A subsequent call to {\bf coconorm} then normalizes these to values from 0 to 255 to produce the array that is ready to be displayed as an RGB image, which is the return value of the function. Other normalization methods as described in e.g.~\cite{lupton2004} are not implemented in this version of COCOPLOT, but can easily be added manually by editing the normalization function {\bf coconorm}. The {\bf cocoplot} function contains additional keywords for:
\begin{itemize}
\item Thresholding the minimum and maximum values in the data cube (values outside these limits are clipped to 0 and 255 in the RGB array).
\item Returning the array with/without displaying it as an image. 
\item Displaying the image a graphics window that is already open.
\item Specifying the desired dimensions of the image that is displayed.
\item Saving a file of the image to a specified location.
\end{itemize}


An animation of individual COCOPLOTs can be created with the {\bf cocovideo} procedure, which takes a 4D data cube (size [$n_x, n_y, n_{\lambda}, n_t$]), filters (size [$n_{\lambda}, 3$]), the frames-per-second rate, and the output file name as required inputs.
The output is a video of COCOPLOTs of the field-of-view $[n_x, n_y]$ with the cube dimension of size $n_{\lambda}$ color-collapsed, for a number of frames $n_t$. 
In addition to the keywords accepted by {\bf cocoplot} described above, {\bf cocovideo} takes keywords for displaying (or not) each image frame in the video as it is being generated and for specifying the beginning and end frames of the video by index of the temporal dimension of the data cube.


More details of these keywords and examples of their use in both {\bf cocoplot} and {\bf cocovideo} are presented in the GitHub code versions.
\bsp	
\label{lastpage}
\end{document}